\documentclass[a4paper,11pt]{article}
\pdfoutput=1 

\usepackage{jinstpub} 

\newcommand\bi{$\mathrm{^{207}Bi}$}
\newcommand\co{$\mathrm{^{60}Co}$}
\usepackage{xcolor}
\usepackage{multirow}
\usepackage{tikz}
\usepackage{float}
\usepackage{subfig}
\def\babar{\mbox{\sl B\hspace{-0.4em} {\small\sl A}\hspace{-0.37em} \sl B\hspace{-0.4em} {\small\sl A\hspace{-0.02em}R}}}

\title{Radiation Hardness of 30 cm Long CsI(Tl) Crystals}

 \author{S. Longo}
 \author{and J.M. Roney}
 \affiliation{University of Victoria\\ 3800 Finnerty Rd, Victoria, BC, Canada. V8P 5C2}

\emailAdd{longos@uvic.ca}

\abstract{
Measurements of the degradation in performance of 30 cm long CsI(Tl) scintillation crystals exposed to 1 MeV photon doses of 2, 10, 35, 100 and 1000 Gy are presented.  The light yield, light yield longitudinal non-uniformity, scintillation decay times, energy resolution and timing resolution of a set of spare crystals from the \babar{} and Belle experiments are studied as a function of these doses.  In addition, a model that describes the plateau observed in the light output loss as a function of dose in terms of increase in concentrations of absorption centres with irradiation is presented.
}


\begin{document}
\maketitle
\flushbottom

\section{Introduction}
\label{}

Thallium doped Cesium Iodide (CsI(Tl)) has found successful application in numerous high intensity $e^+ e^-$ collider experiments such as CLEO \cite{CLEOTDR}, BESIII \cite{BesiiirTDR}, \babar{} \cite{BaBarTDR,BaBarTDR_2} and Belle \cite{BelleTDR}.  Advances in accelerator technology will enable the SuperKEKB $e^+e^-$ collider to reach unprecedented instantaneous luminosities of $8 \times 10^5 \text{ cm}^{-2}\text{s}^{-1}$, which will result in high beam background doses throughout the detector \cite{belle2_tdr}. These projected beam background doses motivate the study of the radiation hardness of large $\sim (5 \times 5 \times 30 \text{ cm}^3)$  CsI(Tl) crystals \cite{Beylin_BelleStudy,Hryn_BabarStudy}.  

The radiation hardness of CsI(Tl) has previously been explored for a range of doses.   The majority of measurements above 100 Gy however have focused on small $\sim (2.54 \times 2.54 \times 2.54 \text{ cm}^3)$ sized crystals \cite{Hamada_EnergyTransProc,Woody_RadCsICsITl,Chowdhury_TolOptAbsBan,Beylin_BelleStudy,Chowdhury_ControlledGrowProc,Zhu_RadDamSciCry}.  By measuring the absorption length, which is defined as the mean free path for self-absorption of optical photons in the crystals at a particular wavelength, results from these studies establish that the light yield degradation of CsI(Tl) from radiation damage is caused by the creation of absorption centres resulting in the decrease of the absorption length of scintillation light in the crystal.  As the path length of scintillation photons is much longer in large sized crystals used in collider experiments, increased self-absorption is expected to have a more significant effect on the degradation of the detected light yield in these crystals.  This was suggested in a past study where a small sample crystal dosed to 500 Gy showed negligible light output loss compared to several 30 cm long crystals dosed to 37 Gy \cite{Beylin_BelleStudy}.   

The purpose of this work is to quantify the radiation hardness of 30 cm long CsI(Tl) crystals at uniform doses up to 1000 Gy using spare crystal samples from the Belle and \babar{} experiments.  With current estimates, in the 10 year lifetime of the Belle II experiment the accumulated dose in crystals in the Belle II endcaps could reach 100 Gy \cite{Boyarintsev_PureCsIRad}.  Our measurement up to 1000 Gy provides an additional factor of 10 safety factor beyond these estimates.  In order to avoid modifications to the crystal wrappings between irradiations the light transmission curves were not measured.  We instead focus on measurements of light output loss, time resolution and non-uniformity as a function of dose as these measurements can be directly used by Belle II collaboration members to evaluate the expected performance of CsI(Tl) crystals in Belle II.  

\section{Crystal Samples}

The CsI(Tl) samples studied consist of six spare crystals from the \babar{} experiment \cite{BaBarTDR} and two spare crystals from the Belle experiment \cite{BelleTDR}. The crystals were grown in the late 1990's and their manufacturer and length are listed in Table \ref{tab:crysamples}. The crystals had a trapezoidal geometry and were tapered such that the readout end was nominally $5 \times 5 \text{ cm}^2$ and the front end was nominally $4 \times 4 \text{ cm}^2$. All crystals were wrapped in 200 $\mu$m thick teflon and thin layers of aluminium and mylar in order to optimize light collection efficiency and crystals were stored in a low humidity environment.   

\babar{} crystals used an R5113-02 Hamamatsu photomultiplier tube (PMT) with photocathode diameter of 46 mm \cite{PMT_datasheet} and an air optical coupling for scintillation light readout.  The Belle crystals used two $10 \times 20 \text{ mm}^2$ Hamamatsu S2744-08 photodiodes glued to the rear of the crystal for light readout \cite{belle2_tdr}. 

\section{Experimental Apparatuses}

\subsection{Uniformity Apparatus}

\begin{figure}[h]
\centering
\includegraphics[width=1\textwidth]{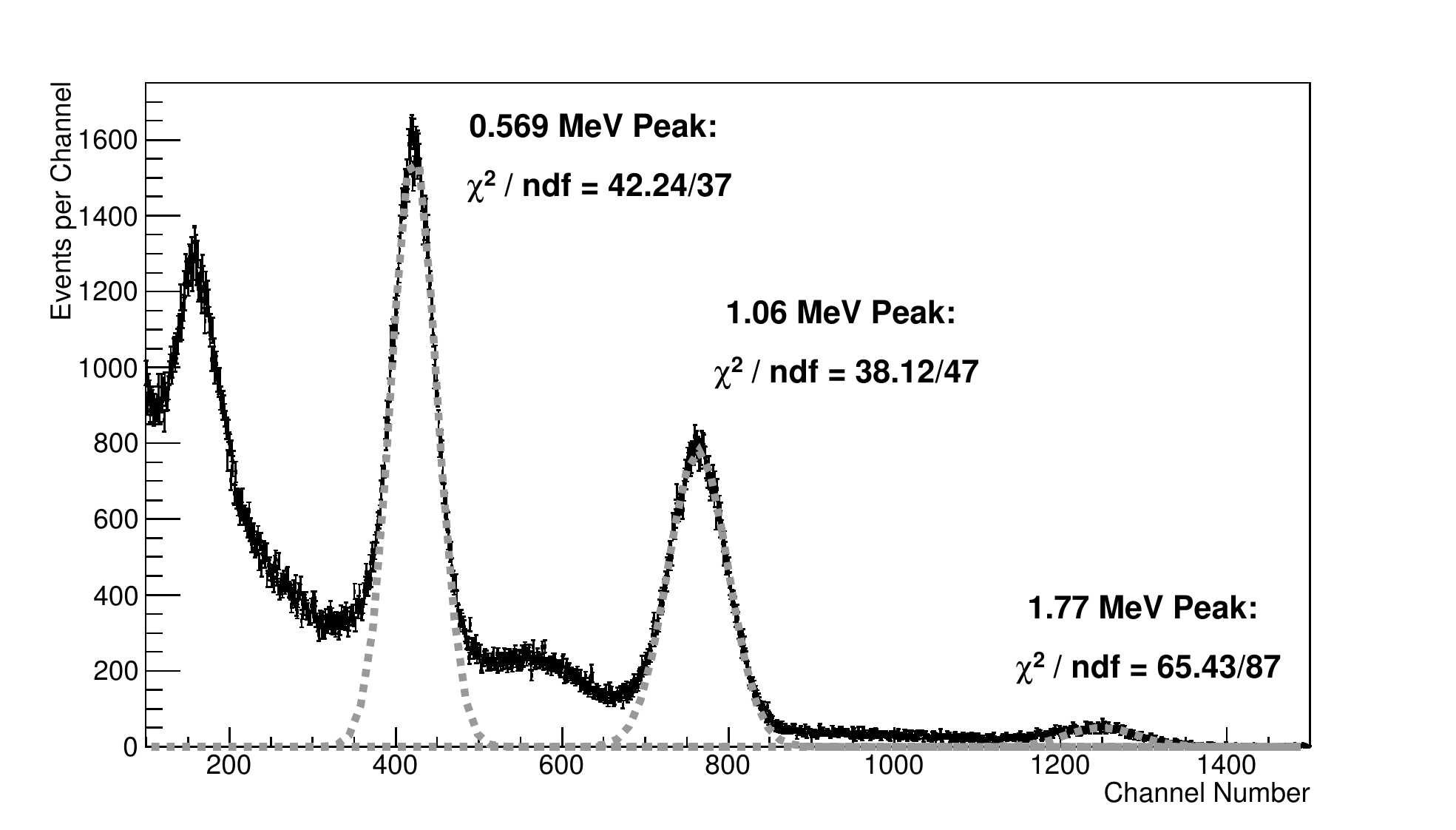}
\caption{Sample \bi{} gamma spectrum measured using \babar{} crystal.  Gaussian fits used to calculate peak channel numbers are overlaid.} 
\label{fig:bispectrum} 
\end{figure}

Light yield measurements for \babar{} crystals were completed using an apparatus consisting of mounts for the CsI(Tl) crystal and PMT as well as a 2.54 cm thick lead collimator placed on a stepper motor controlled track placed along the crystal length.   Light yield measurements were completed using 0.5, 1 and 1.7 MeV gammas from a \bi{} source.  A sample \bi{} spectrum is shown in Figure \ref{fig:bispectrum}.  Each light yield measurement acquired nine \bi{} spectra at 3 cm spacings along the crystal length, beginning 3 cm from the PMT readout end.  During light yield measurements the apparatus was placed in a dark box where temperature and humidity were monitored.    

The signal chain for the \babar{} crystals consisted of a PMT connected to a Tennelec TC 241 Amplifier with 6 $\mu$s shaping time. The pulse heights of the shaped pulses were then recorded.  For the duration of the study an undosed reference crystal was used to monitor the stability of the electronics.  Scintillation pulses from cosmic energy deposits were also recorded at each irradiation stage by self-triggering on the cosmic events.

\begin{table}[h]
\centering
\caption{Crystal samples studied indicating detector origin and manufacture.  *Used as reference crystal.} \label{tab:crysamples} 
\smallskip
\begin{tabular}{|c|c|c|c|c|}
\hline
     Detector Origin           &     Light Readout       & Crystal ID & Manufacturer &  Length  (cm) \\ \hline
\multirow{3}{*}{Belle}      & \multirow{3}{*}{2 PIN Diodes} & 320017   &    &\multirow{7}{*}{30}  \\ \cline{3-3} 
                                       &                                              & 334017   &   Shanghai Institute    &  \\ \cline{3-3} 
                                       &                                               & 315065* &           of Ceramics               &    \\ \cline{1-3} 
   \multirow{7}{*}{\babar{}} & \multirow{7}{*}{ Air optical coupling} & BCAL02410*  &          &   \\ \cline{3-3} 
                                        &                                            & BCAL02676   &                          &    \\ \cline{3-4} 
                                        &                                            & BCAL03348   &  \multirow{3}{*}{Crismatec}   &   \\ \cline{3-3} 
                                        &                                            & BCAL02234   &                          & \\ \cline{3-3} 
                                        &                                            & BCAL03334   &                          &    \\ \cline{3-5} 
                                        &                                            & BCAL05881   &Kharkov Institute  & \multirow{2}{*}{31.5}\\ \cline{3-3} 
                                        &                                            & BCAL05883   &    for Single Crystals &   \\ \hline
\end{tabular}
\end{table}

\subsection{Belle Crystal Measurement Apparatus}

The readout electronics for the Belle crystals consisted of two PIN diodes connected to a Belle pre-amp which integrated and shaped the pulses with a 1 $\mu s$ shaping time \cite{BelleTDR}.  Each shaped diode pulse was processed in a readout board developed at the University of Victoria that replicated the electronics to be used at Belle II \cite{Longo_Measurements}.  For light yield measurements, the diode pulses were summed using the readout board then digitized using a Tektronix oscilloscope.  For time resolution measurements, the individual shaped diode pulses were both digitized.

The diodes glued to the back of the Belle crystals did not have sufficient sensitivity to detect the low energy gammas from the \bi{} source, therefore a Cosmic Ray Test Stand (CRTS) was assembled to perform light yield uniformity measurements on the Belle crystals utilizing cosmic ray energy deposits.  The CRTS used seven plastic scintillator paddles arranged in three horizontal planes with the sample crystal placed between the top two planes.  The six paddles in the top two planes were used to section the Belle crystals longitudinally into three 10 cm long areas corresponding to near diode, middle and far from diode positions allowing for light yield uniformity measurements to be made.  The readout system of the Belle crystals was triggered when three vertically aligned paddles of the CRTS were simultaneously triggered.   A 2.54 cm thick lead plate was placed between the bottom two planes of the CRTS in order to select on minimum ionizing particles.  In order to monitor the stability of the apparatus a reference crystal was measured simultaneously to a irradiated crystals for all light yield measurements.  

\begin{figure}[h]
\centering

\subfloat[]{\includegraphics[width=0.50\textwidth]{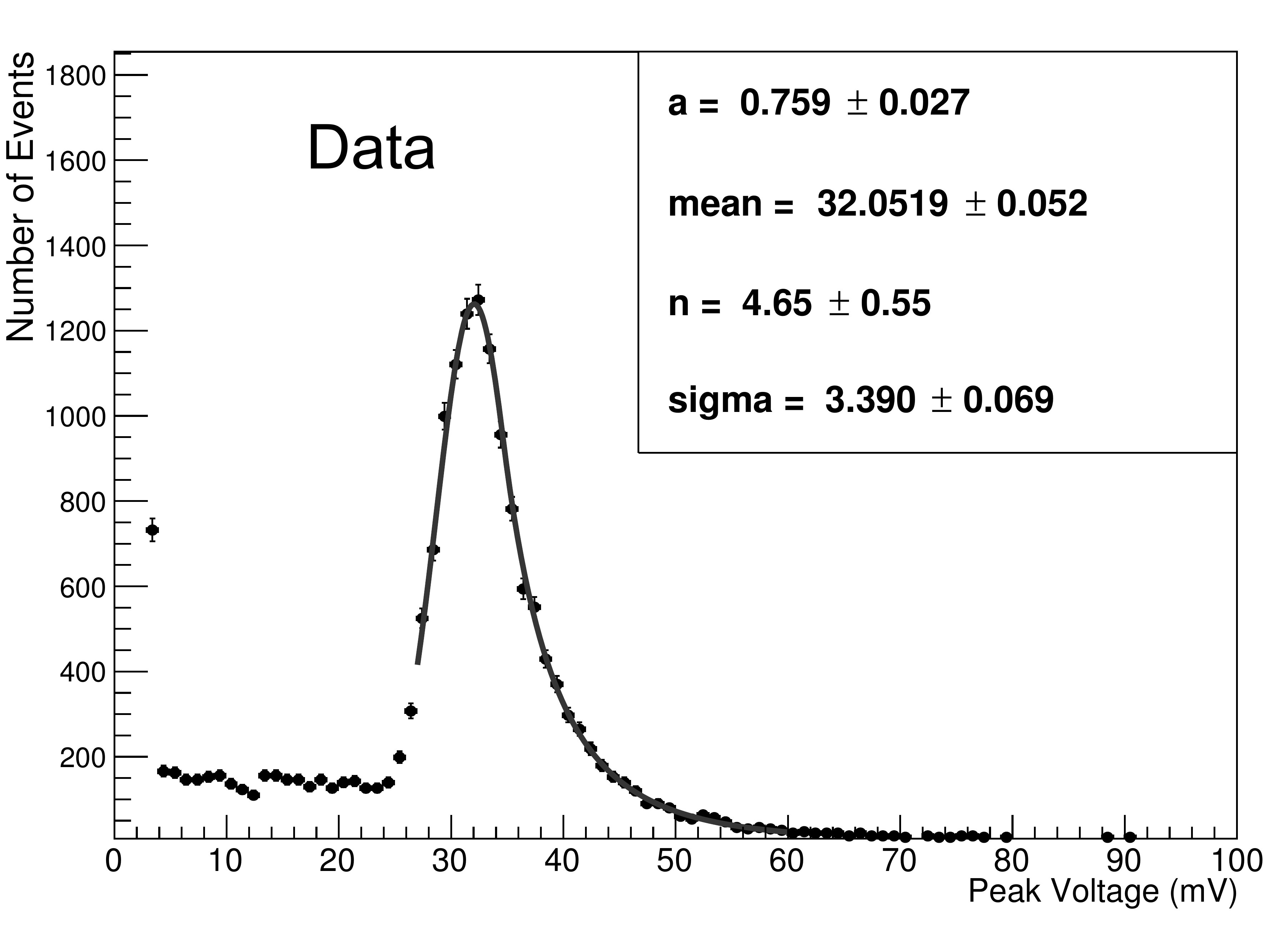}\label{fig:bellephh} } 
\subfloat[]{\includegraphics[width=0.50\textwidth]{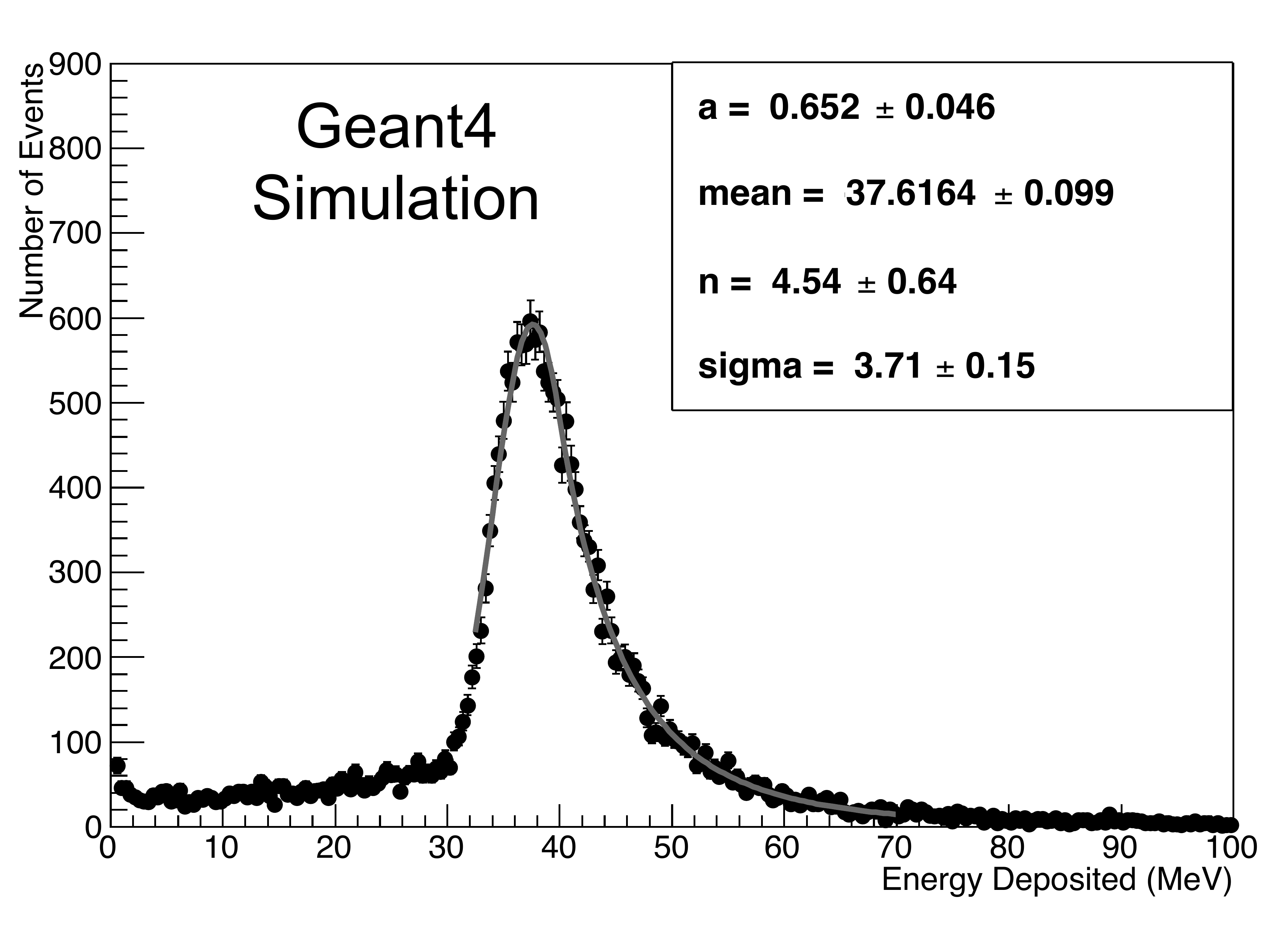}\label{fig:bellegeant}}

\caption{a) Sample pulse height histogram recorded using Belle crystal and CRTS. The fit to a reversed CBF is shown. b) Simulated results of energy deposits from muons using the CRTS.} 
\end{figure}

A sample pulse height histogram recorded from a cosmic run using one of the Belle crystals is shown in Figure \ref{fig:bellephh}.  Overlaid is a sample fit for a reversed Crystal Ball Function (CBF) \cite{roofit_ref} where the mean corresponds to the mean of the Gaussian component of the CBF. This was used to quantify the light yield of the crystals.  The CRTS was simulated using Geant4 \cite{geant4} in order to quantify the distribution of energy deposited in the crystals from muons \cite{Longo_Measurements}. The distribution of energy deposited for muons satisfying the trigger conditions of the CRTS is plotted in Figure \ref{fig:bellegeant}.

\subsection{CsI(Tl) Irradiation Methods}
\label{sec_doseinfo}

Crystals were irradiated using $\sim 1$ MeV gammas from a \co{} source at the National Research Council of Canada (NRC) Measurement Science and Standards facility.   Crystals were dosed by the NRC to an accuracy of $\pm 2\%$ and all dose measurements and calculations are detailed in reference \cite{NRC_Dose_Report}. The average dose rate for the irradiations was approximately $0.2 \text{ Gy}/ \text{min}$. All crystals except \babar{} crystal 3334 were given uniform doses across the crystal width by rotating the crystals 180 degrees half way between irradiations.  Crystal 3334 was dosed such that the non-readout end faced the \co{} source, resulting in an exponentially attenuated dose along the crystal's longitudinal length.  Doses quoted for this crystal are for the dose in the first fifth of the crystal length.  

\section{Results}

\subsection{Light Yield Degradation}

\begin{figure}[H]
\centering
\includegraphics[width=0.99\textwidth]{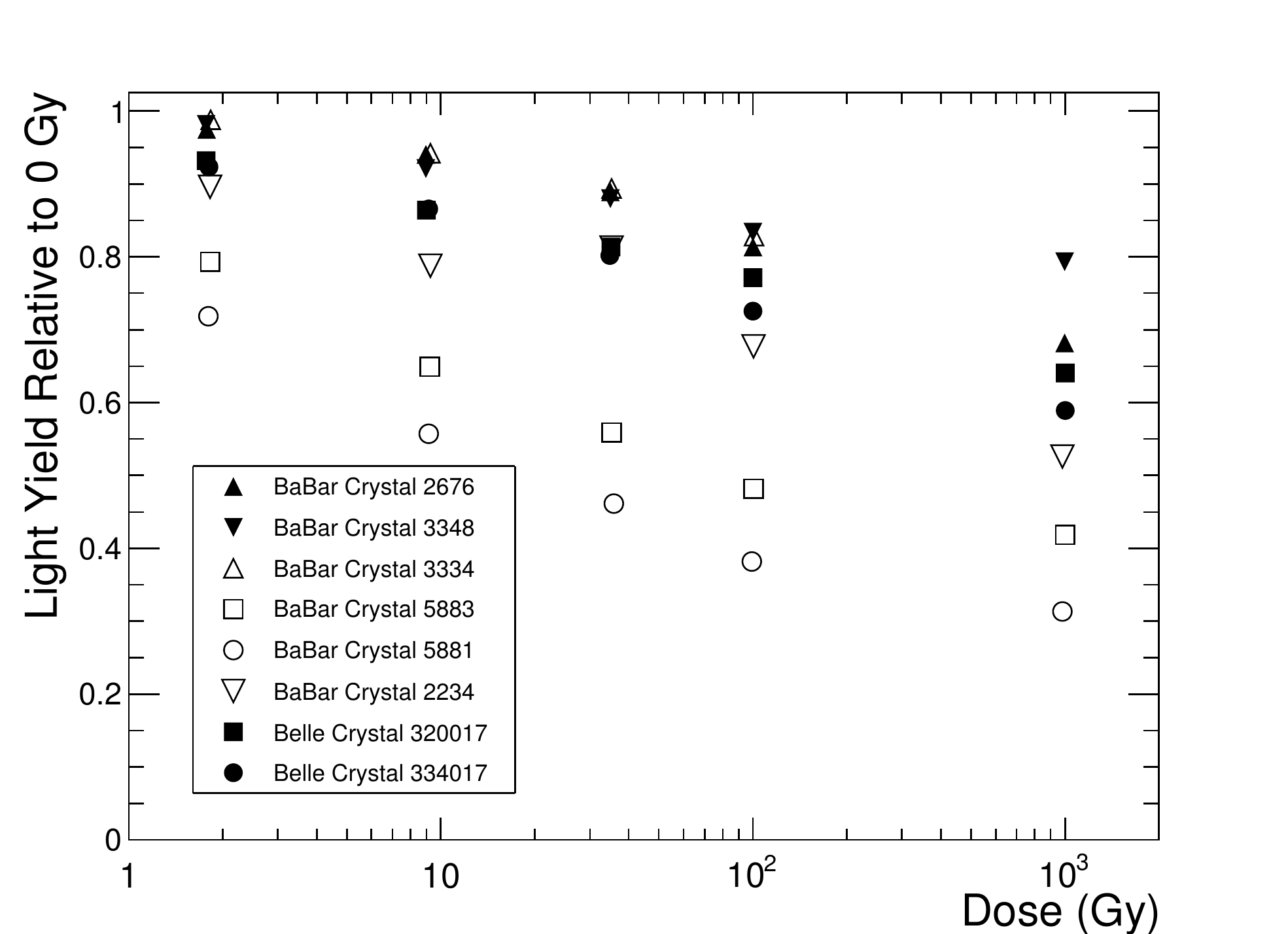}
\caption{Light yield relative to 0 Gy for all samples studied.  Combined statistical and systematic error bars are smaller then the data points.  Total error on dose and light yield measurements are approximately $\pm 2\%$ and $\pm 0.5\%$, respectively. } 
\label{fig:AllLyvsDose} 
\end{figure}

The light yield relative to 0 Gy dose for all crystals at all irradiation stages is shown in Figure \ref{fig:AllLyvsDose}.  For the spare Belle crystals, our measurements agree with a previous study of uniformly irradiated Belle CsI(Tl) crystals up to 37 Gy \cite{Beylin_BelleStudy} and extend these results up to 1000 Gy showing that the light yield degradation continues to plateau requiring an order of magnitude more dose to have the same relative drop in light yield.  This plateauing light output loss trend is observed for both \babar{} and Belle crystals.

To validate whether the controlled uniform irradiation method using \co{} is sufficient in replicating beam background doses present during operation at $e^+e^-$ colliders, a comparison of our measurements can be made to the light output loss observed in the Belle detector crystals over their 10 year lifetime.  The accumulated dose in the Belle calorimeter crystals was measured to be 1-4.5 Gy and an average light output loss of 8\%  was observed \cite{belle2_tdr}. This is consistent with the 2 Gy irradiation stage of this study as the light output loss of both Belle crystals was measured to be 8\%. 

For the all crystals studied it was found that the magnitude of the light output loss was correlated to the shipment batch of each crystal when originally received from the manufacturer which we assume is related to the production lot.  Specifically crystals 5881 and 5883 from Kharkov Institute for Single Crystals both lost over 20\% of their 0 Gy light yield after 2 Gy and were the least radiation hard.  The two Belle crystals also show similar light output loss up to 1000 Gy.  The two uniformly irradiated \babar{} crystals from Crismatec, 3348 and 2234, had varying radiation hardnesses however these two crystals originated from different Crismatec batches.    These correlations suggest that crystals grown in identical conditions can be expected to have similar radiation hardnesses and supports theories suggesting the radiation hardness of CsI(Tl) is dependent on the growth procedure \cite{Zhu_RadDamSciCry}.

\begin{figure}[H]
\centering
\includegraphics[width=1\textwidth]{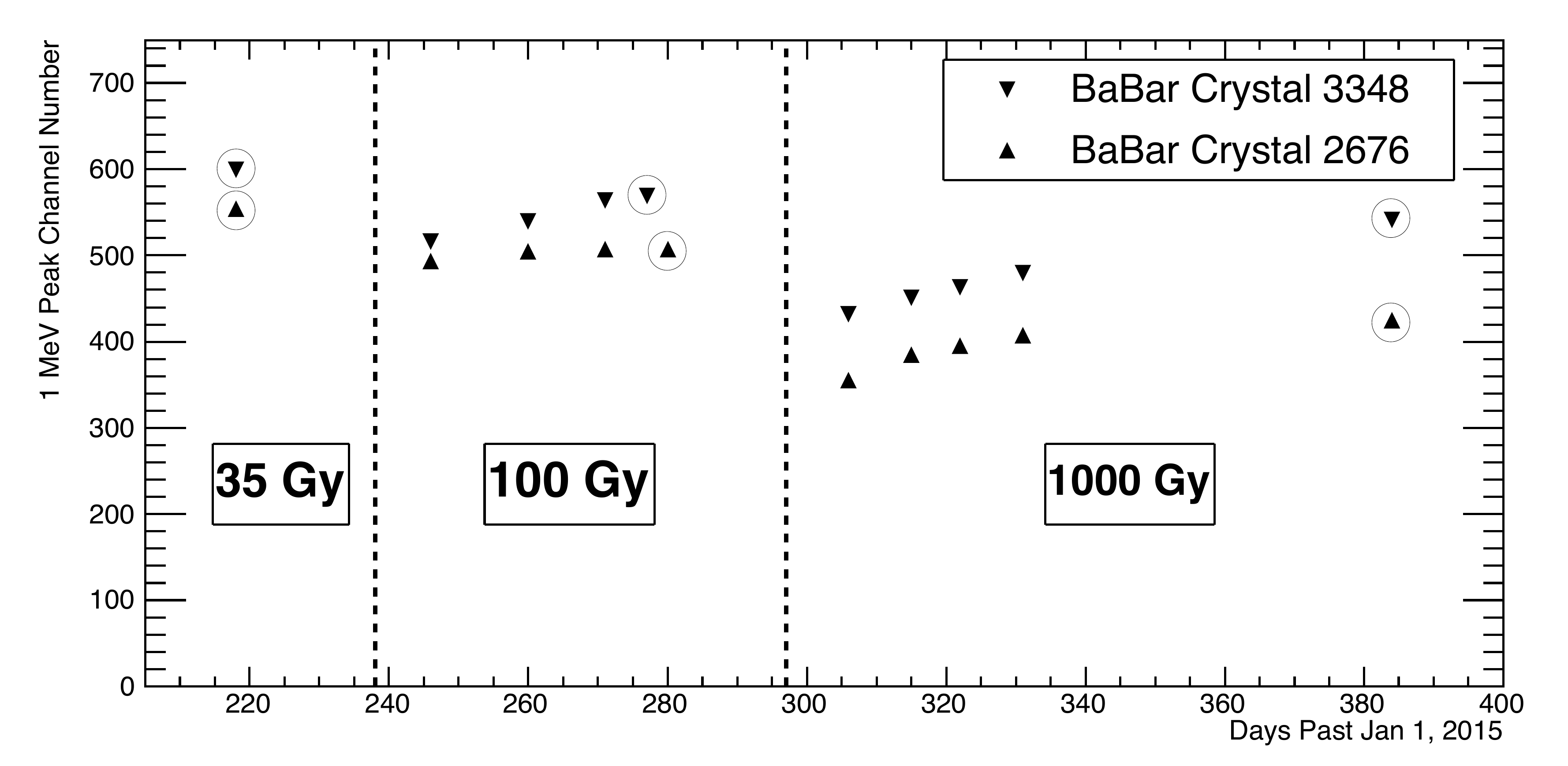}
\caption{Long term recovery observed in two \babar{} crystals after 100 Gy and 1000 Gy doses. Dashed lines indicate the day the crystal doses were completed. Circled points indicate recovered values used for light output loss calculation.} 
\label{fig:recovery_plots} 
\end{figure}

\subsection{Crystal Recovery}

At the 35, 100 and 1000 Gy irradiation stages the light yield of the \babar{} crystal samples was observed to initially have a large decrease then undergo a recovery phase.  Shown in Figure \ref{fig:recovery_plots} for two \babar{} crystals, the recovery period was observed to be several weeks before the light yield stabilized.  Due to the long term light yield recovery observed at these stages, the light yield measurements reported in Figure \ref{fig:AllLyvsDose} for the 35, 100 and 1000 Gy irradiation stages are of measurements made more than five weeks after irradiation.   For the Belle crystals, no recovery was observed.  Similar recovery in irradiated CsI(Tl) crystals has also been reported in past radiation hardness studies \cite{Woody_RadCsICsITl,Hamada_EnergyTransProc,Zhu_RadDamSciCry}.

No significant long term recovery was observed for the majority of \babar{} crystals at the 2 and 10 Gy irradiation stages during the two week period after irradiation.  The light yield of crystal 2234 was measured to increase from the 10 Gy to 35 Gy irradiation stage however, this increase in light yield was measured after a two month recovery period.  Measured two weeks after the 35 Gy irradiation, the light yield of crystal 2234 was lower than the 10 Gy stage.  This suggests that crystal 2234 was undergoing slow recovery at the 10 Gy irradiation stage and the increase at 35 Gy is due to this crystal not reaching a long term recovery value at 10 Gy.   

\begin{figure}[H]
\centering
\includegraphics[width=0.75\textwidth]{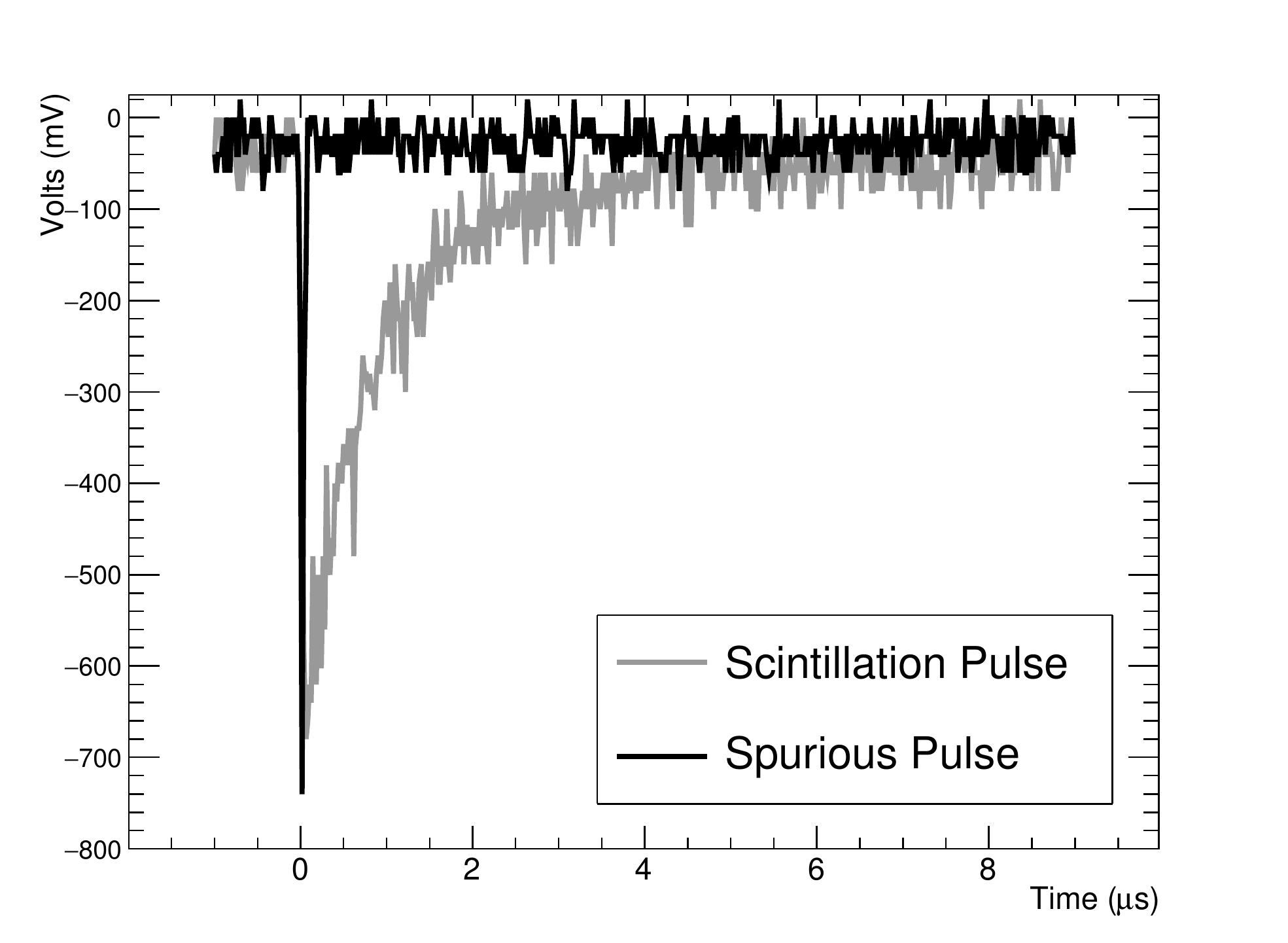}
\caption{Comparison of a scintillation pulse and spurious pulse.} 
\label{fig:pulse_overlay} 
\end{figure} 

\subsection{Observation of Non-scintillation Pulses in \babar{} Crystals}

Spurious pulses were observed to be present in irradiated \babar{} crystals at the 100 Gy and 1000 Gy irradiation stages.  An overlay of a typical scintillation pulse from a cosmic event and spurious pulse is shown in Figure \ref{fig:pulse_overlay}.  This is a reproducible effect that is only present when the PMT is attached to the crystals and is present for all \babar{} crystals. The frequency of the spurious pulses was observed to be several Hertz two weeks after irradiation (measurements prior to two weeks were not made).  The pulses were observed to be similar in peak amplitude to the CsI(Tl) scintillation pulses from cosmic deposits ($\sim 1$ V) however the spurious pulses had a much shorter time scale thus the integrated charge was much smaller than a cosmic pulse. This allowed for discrimination of these pulses during analysis however when acquiring data by self-triggering on pulse heights these pulses created dead times.  As the \babar{} crystals underwent their recovery period the frequency of these pulses was found to decrease.

This effect was not observed in the Belle crystals although we note that the output pulses from the  Belle crystals were of the integrated diode charge.  The origin of these pulses is unknown and has not been reported in previous CsI(Tl) radiation hardness studies.  We note that the PMT used with the \babar{} crystals had a UV window with spectral range from 185-650 nm peaking at 420 nm \cite{PMT_datasheet} and the diodes used with the Belle crystals had a spectral range from 340-1100 nm, peaking at 960 nm \cite{Diode_datasheet} .  From this information it is possible that the spurious pulses were not observed in the Belle crystals because the Belle crystal signal was integrated or the wavelength of the source is outside the spectral window of the diodes used with the Belle crystals, or a combination of both effects. We report this as an observation that may be explored further in future studies.

\begin{figure}[H]
\centering
\includegraphics[width=0.75\textwidth]{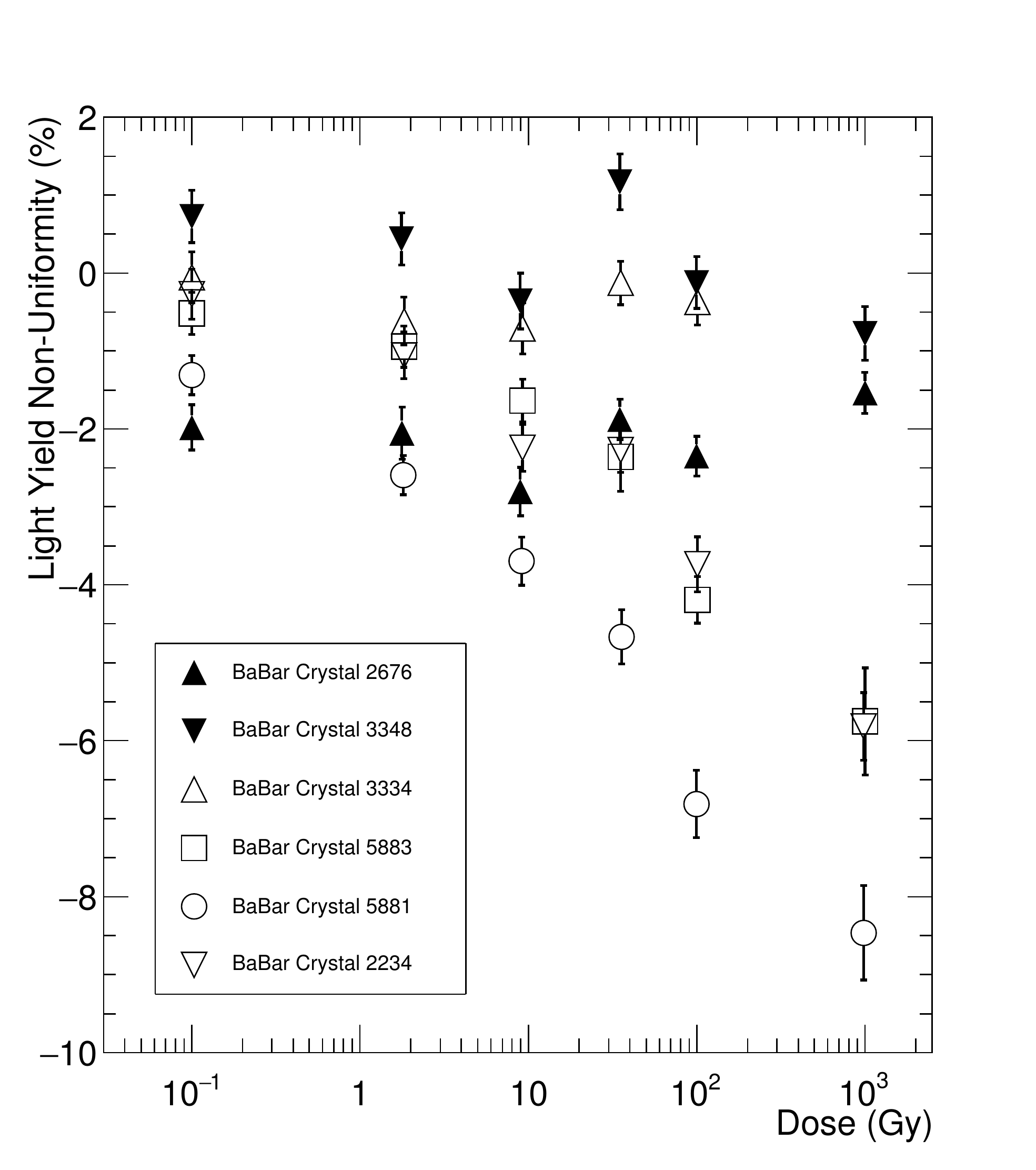}
\caption{Longitudinal light yield non-uniformity of \babar{} crystals as a function of dose. Due to logarithmic scale 0 Gy dose value is shown at $10^{-1}$ Gy.} 
\label{fig:univsdose} 
\end{figure}

\subsection{\babar{} Crystal Changes in Longitudinal Light Yield Non-Uniformity}

The longitudinal light yield non-uniformity defined by the difference in the light yield at a distances of 3 cm and 27 cm from the PMT normalized to the average light yield of the nine measurements along the crystal length is plotted in Figure \ref{fig:univsdose} for all \babar{} crystals.  Errors presented in Figure \ref{fig:univsdose} are statistical errors only calculated from propagating peak mean error from Gaussian fit and pedestal mean error.  For crystals 3348, 2676 and 3334 which had no significant changes in non-uniformity, fluctuations about the mean non-uniformity values is observed within a few statistical sigma.  As this measurement is calculated from a ratio many systematic effects cancel.  A systematic error for this measurement could arise from misalignment of the collimator however, it is estimated that the maximum change in non-uniformity from this effect is less then 0.1\% of the measured values.

From Figure \ref{fig:univsdose}, it is observed that crystals 2234, 5881 and 5883, which had poor radiation hardness, also had large changes in non-uniformity.  The changes in longitudinal light yield non-uniformity observed for these crystals were such that the non-readout end of the crystals had the largest drops in light yield.   This trend is expected from increased self-absorption throughout the crystal \cite{Hryn_BabarStudy}.  For crystals 2676 and 3348 which had good radiation hardness, less than 2\% change in uniformity is observed as the dose increased to 1000 Gy.

\subsection{Belle Crystal Changes in Longitudinal  Light Yield Non-Uniformity}

\begin{figure}[H]
\centering
\subfloat[]{\includegraphics[width=0.7\textwidth]{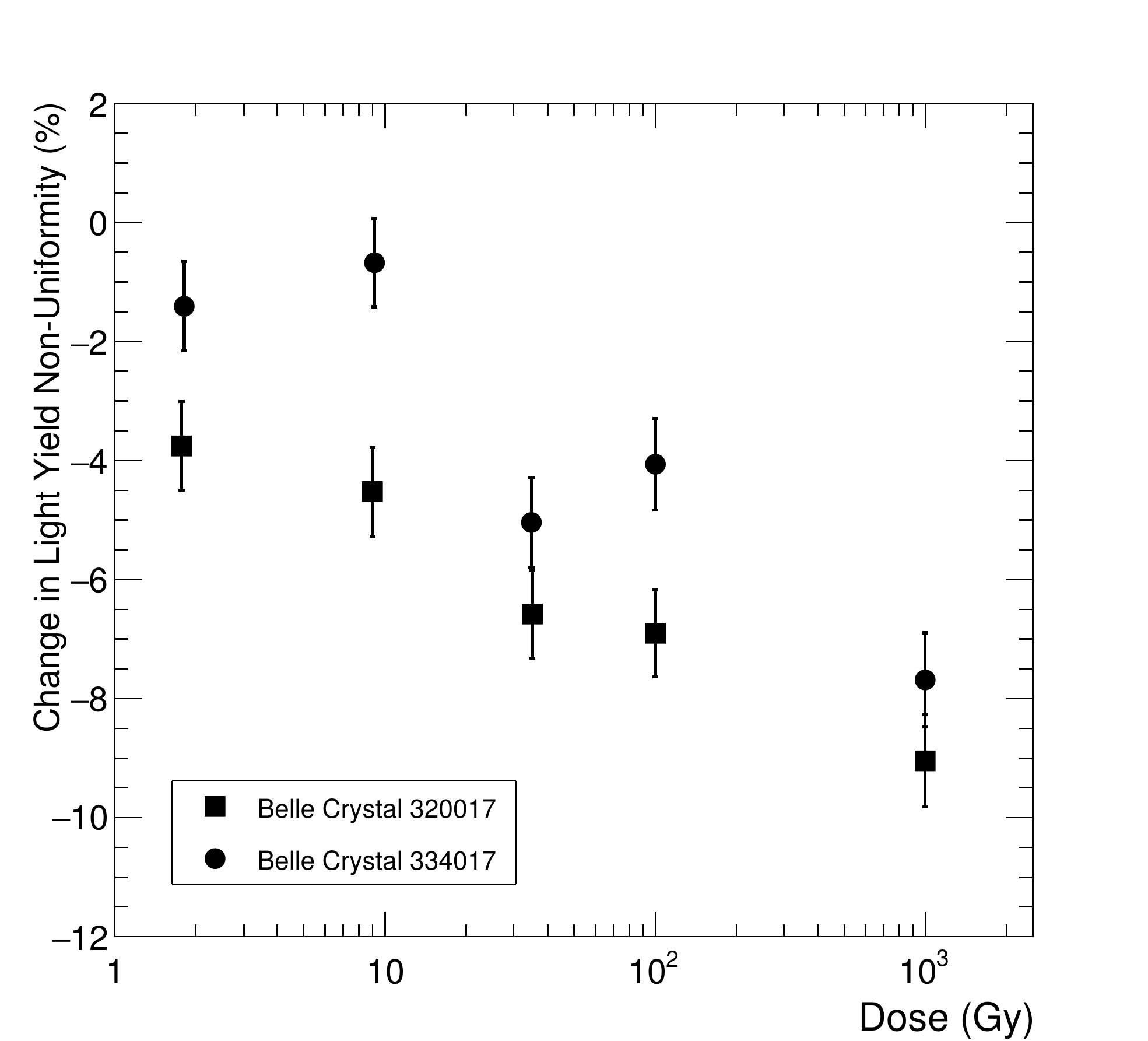}\label{fig:bellechangenuf} } 

\caption{Change in longitudinal light yield non-uniformity observed in Belle crystals.} 
\end{figure}

As the trigger logic of the CRTS required the particle depositing energy to traverse through the full crystal height, the path length for a particle satisfying the CRTS trigger conditions is larger near the diode end due to the crystal taper. As a result the absolute non-uniformity of the Belle crystals could not be calculated due to the crystal taper creating non-uniform cosmic energy deposits along the crystal length.   Instead the change in longitudinal light yield uniformity relative of 0 Gy dose was determined by calculating the difference in the light yield relative to 0 Gy for CRTS trigger locations far from diode and near diode.  This is shown in Figure \ref{fig:bellechangenuf} where is it observed that large changes in non-uniformity were present in the Belle crystals.  As with the \babar{} crystals, the light yield far from the readout end was found to degrade faster suggesting a decrease in the absorption length for scintillation photons in the crystals.

\subsection{Longitudinally Non-Uniform Irradiated Crystal}

\babar{} crystal 3334 was irradiated such that the non-readout end of the crystal was directed at the \co{} source resulting in an exponentially attenuated dose along the crystal length and, as mentioned in Section \ref{sec_doseinfo}, the dose calculated for this crystal is for the first fifth of the crystal length.   In order to compare the uniform irradiation to the non-uniform irradiation method, the radiation hardness properties of crystal 3334 can be directly compared to crystal 3348 as these crystals were from the same crystal batch from Crismatec.

\begin{figure}[H]
\centering
\includegraphics[width=0.7\textwidth]{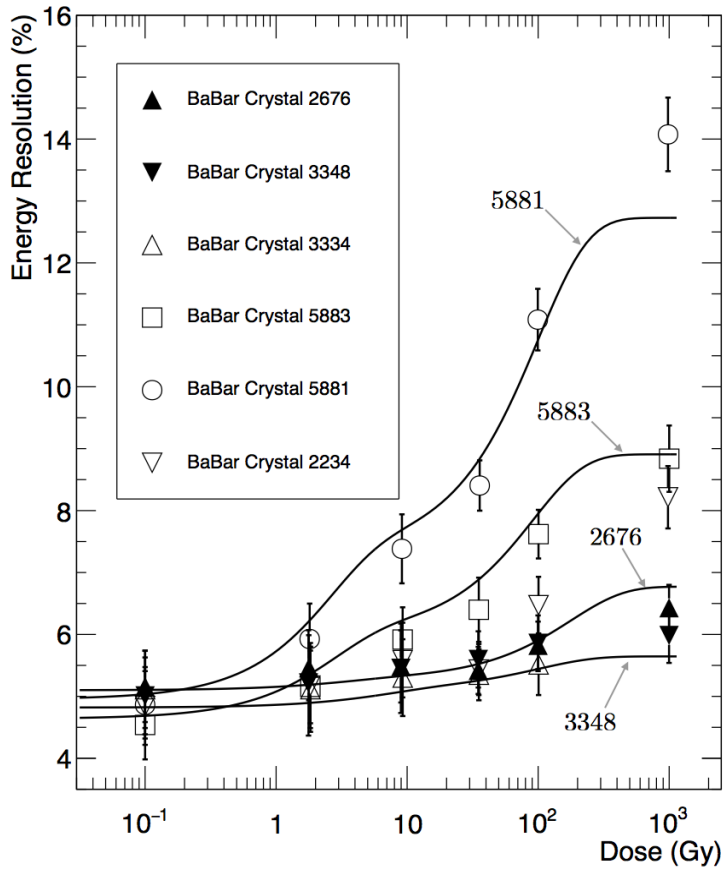}
\caption{Energy resolution of 1.06 MeV photons measured with the \babar{} crystals at each irradiation stage. Predicted energy resolution degradation curves for crystals 2676, 3348, 5881 and 5883 are overlaid. The shapes of the overlaid curves reflect the parameterized light output loss fits of Equation \ref{eqn_model} (shown in Figure \ref{fig:modelfitindependent}), which describes the saturation effects in a model with two defect types discussed in Section \ref{ModelFitsSection}.  Due to logarithmic scale 0 Gy dose value is shown at $10^{-1}$ Gy.} 
\label{fig:energyres_plot} 
\end{figure} 

Comparing samples 3348 and 3334 it is observed that both crystals had very similar radiation hardness properties.  The overall light output loss shown in Figure \ref{fig:AllLyvsDose} is observed to be similar up to 100 Gy, with both crystals showing good radiation hardness.  The longitudinal light yield uniformity of these crystals shown in Figure \ref{fig:univsdose} was observed to not change up to 100 Gy.  This shows that for 30 cm long crystals there is not a significant difference observed between the two irradiation methods and that these studies are valid for different irradiating photon directions.

\subsection{Changes Crystal Colour}

The \babar{} crystals were observed to change colour from clear to red/pink. The degree of colour change was observed to be correlated with the light output of the crystal such that crystals which were very degraded also had large changes in colour.  Images of the crystal colours can be found on page 79 in reference \cite{Longo_Measurements}. 

\begin{figure}[H]
\centering
\includegraphics[width=0.85\textwidth]{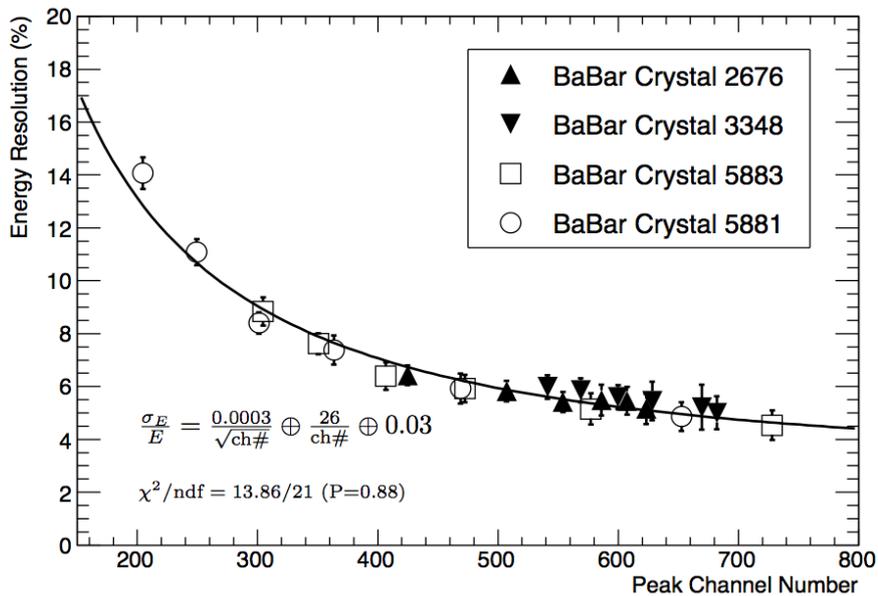}
\caption{Energy resolution of 1.06 MeV photons as a function of light yield measured with the \babar{} crystals.  Fit to Equation \ref{eqn_EresEqn} is overlaid with results of fit shown.} 
\label{fig:energyresvsLY_plot} 
\end{figure}

\subsection{Energy Resolution Degradation in \babar{} crystals}

The energy resolution for 1.06 MeV \bi{} photons is shown in Figure \ref{fig:energyres_plot} measured for the \babar{} crystals at each irradiation stage.  The degradation in energy resolution observed was found to be correlated to the light output loss observed and changes in non-uniformity observed such that crystals with large drops in light output and increases in non-uniformity also had large degradation in energy resolution.  

The magnitude of the energy resolution degradation observed however was larger than expected assuming the degradation is only from a reduction in photoelectron statistics.  The measured energy resolution as a function of measured light yield, is plotted in Figure \ref{fig:energyresvsLY_plot} using data from \babar{} crystals.  The energy resolution as a function of energy can be described by Equation \ref{eqn_EresEqn} \cite{Groupen_particledetectors},

 \begin{equation}
 \label{eqn_EresEqn}
\frac{\sigma_E}{E} = \frac{a}{\sqrt{E}} \oplus \frac{b}{E} \oplus c
 \end{equation}

 where $a$ is the stochastic term from photoelectron statistics, $b$ is the noise term from electronics, $c$ is a constant term and $\oplus$ indicates addition in quadrature \cite{Groupen_particledetectors}.  The fit result of the data to this function is shown in Figure \ref{fig:energyresvsLY_plot}.  This result characterizes the energy resolution of the Uniformity Apparatus including contributions from the \babar{} crystals and the signal chain electronics.  The large $b$ term in the fit results indicate that the degradation in energy resolution observed is dominated by noise from the electronics used in the signal chain.  By using electronics optimized for low noise,  the magnitude of degradation in energy resolution can likely be reduced.

\begin{figure}[H]
\centering

\subfloat[]{\includegraphics[width=0.8\textwidth]{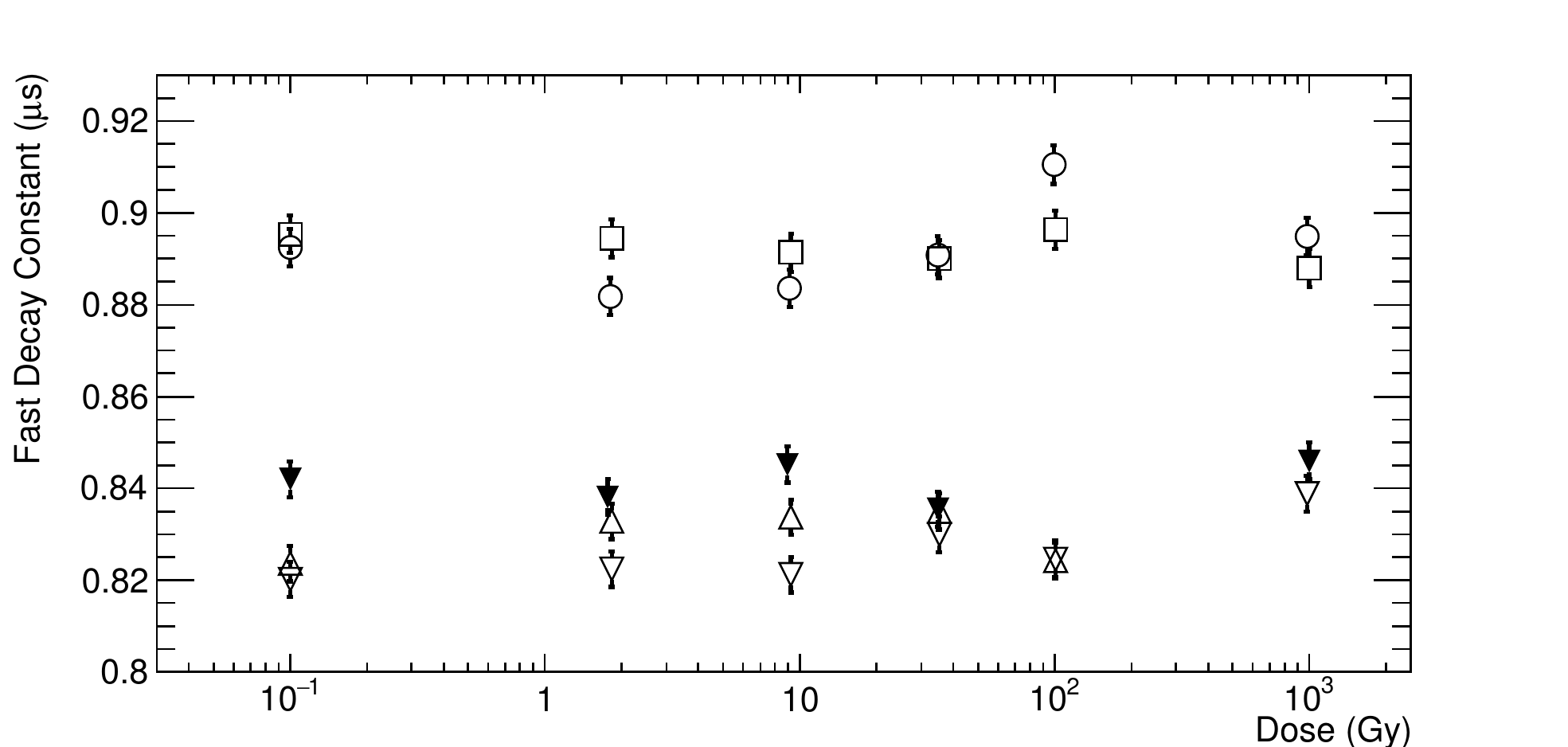}\label{fig:fast} } 

\subfloat[]{\includegraphics[width=0.8\textwidth]{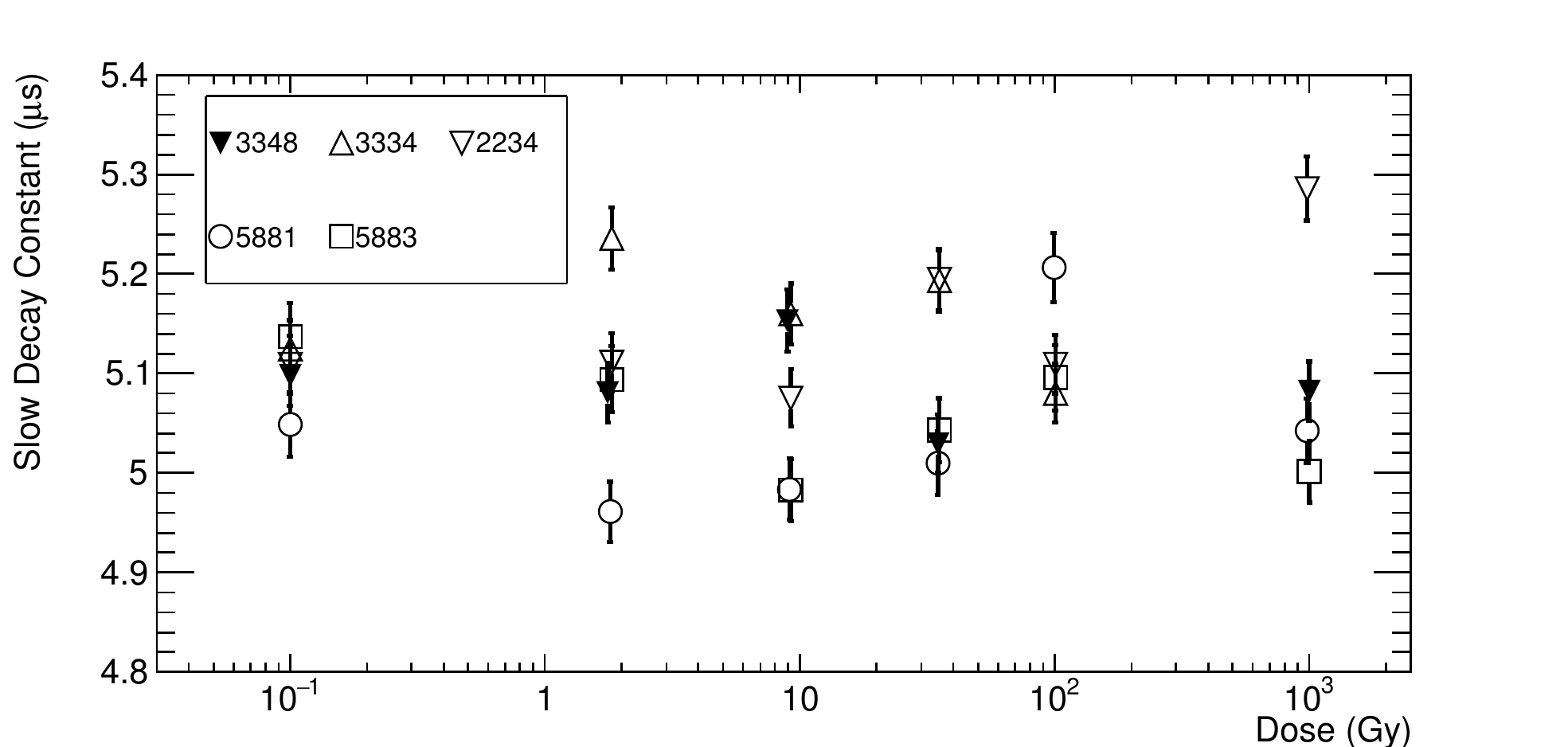}\label{fig:slow} } 

\caption{a) Fast and  b) Slow decay constants of \babar{} crystals as a function of dose.  Note 100 Gy measurement for crystal 3348 could not be completed due to large dead time from spurious pulses mentioned in the text. Due to logarithmic scale 0 Gy dose value is shown at $10^{-1}$ Gy.} 
\label{fig:fastau} 
\end{figure}

Using the fit results for Equation \ref{eqn_EresEqn} shown in Figure \ref{fig:energyresvsLY_plot} and the analytic curves for the light yield as a function of dose shown in Section \ref{ModelFitsSection} on Figure \ref{fig:modelfitindependent}, the predicted energy resolution degradation as a function of dose curves, given the light output loss observed, is overlaid on Figure \ref{fig:energyres_plot}.  From this overlay good agreement is observed between measured and expected energy resolution degradation.  In addition the predicted energy resolution degradation plateaus at large dose due to the plateau in light output loss, as expected.

\subsection{Scintillation Decay Times}

The scintillation decay times of the \babar{} crystals was calculated by averaging over several hundred normalized cosmic pulses then fitting two decay constants to the averaged pulse in the region of 0 - 9 $\mu$s.  The fast and slow decay constants for several \babar{} crystals calculated using this method are shown in Figures \ref{fig:fast} and \ref{fig:slow} as a function of dose and are found to be consistent with previous measurements of the scintillation decay time of CsI(Tl) at room temperature \cite{Valentine_TempCsITL}. 

No changes in pulse shape is observed suggesting the scintillation mechanism was unaffected from the irradiations.  However the fast decay times for crystals 5881 and 5883 which both originated from the same manufacturer are measured to be approximately 0.06 $\mu$s larger than the remaining crystals studied. From Figure \ref{fig:AllLyvsDose}, crystals 5881 and 5883 were also observed to have relativity large light output loss compared to other crystals studied at early dose stages.   Further studies of the correlation between the fast decay time and the radiation hardness of CsI(Tl) crystals could determine if this is a more general phenomenon. 

\begin{figure}[H]
\centering
\includegraphics[width=0.8\textwidth]{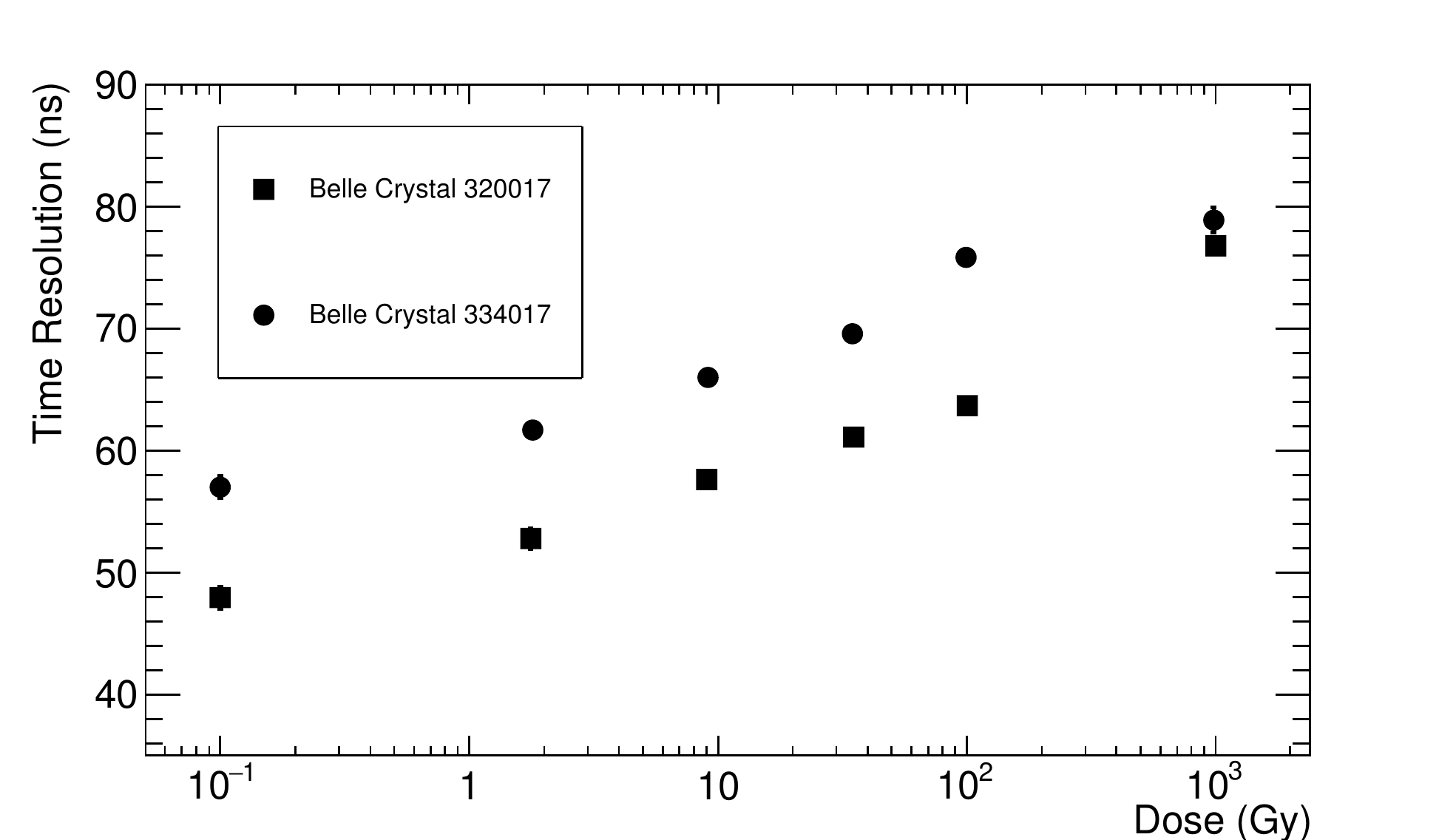}
\caption{Timing resolution of Belle crystals for a $\sim$ 20 MeV energy deposit measured as a function of dose.  Due to logarithmic scale 0 Gy dose value is show at $10^{-1}$ Gy.} 
\label{fig:TR_plot} 
\end{figure} 

\subsection{Time Resolution of Belle Crystals}

Using constant fraction discrimination to determine pulse timing, the time resolution of the Belle crystals was calculated by measuring the standard deviation of the time difference between the pulses from each diode on the crystal using cosmic events and then dividing the standard deviation of the distribution by $\sqrt{2}$ in order to account for the contributions from each diode.  This is plotted in Figure \ref{fig:TR_plot} for both Belle crystals. As we are not summing the diode pulses for this measurement, the number of photoelectrons per energy deposit available is halved compared to a summed pulse.  Therefore the time resolution measurement using the individual diodes is equivalent to measuring the time resolution for a summed pulse with energy deposits of 15-25 MeV which is half of the energies shown in Figure \ref{fig:bellegeant}.
 
The time resolution of Belle crystals is reported in the Belle II Technical Design Report to be $\sim$60 ns at 10 MeV and $\sim$22 ns at 30 MeV \cite{belle2_tdr} and is consistent with our 0 Gy measurement. The constant offset between the two crystals is due to a small difference in the thickness of the crystals leading to higher cosmic energy deposits in crystal 320017.  As the scintillation decay times of the \babar{} crystals was observed to be unchanged for all irradiation stages, the degradation of timing resolution observed is likely a result of the light output loss in the Belle crystals.

\section{Modelling Light Output Loss in Crystals with Increased Absorption}

Assuming the light yield degradation is a result of impurities or imperfections forming absorption centres, the light yield relative to 0 Gy, $ \text{LY}(D)$, as a function of dose, $D$, is given by the Beer-Lambert Law in Equation  \ref{eqn_rel_LY} where $L$ is the average path length for a scintillation photon in the crystal and $A_i$ is the absorption length from the $i^\text{th}$ defect type.  

 \begin{equation}
 \label{eqn_rel_LY}
 \text{LY}(D) = \exp \Big( -L \sum_i \frac{1}{A_i(D)} \Big)
 \end{equation}
 
 As the number of absorption centres per unit volume in the crystal, $N_i(D)$, increases the absorption length of the crystal will decrease.  This is described by Equation \ref{eqn_A_prop} where $\kappa$ is a constant of proportionality relating number of absorption centres to absorption length.
 
  \begin{equation}
 \label{eqn_A_prop}
 A_i (D) = \frac{\kappa}{N_i(D)}
 \end{equation}
 
It is assumed the creation of an absorption centre from a defect of type $i$ is a process described by a parameter $\rho_i$ which has units of Gray.  This parameter characterizes the mean amount of radiation needed to convert a defect into an absorption centre of type $i$.  The number of absorption centres for a specific defect type is limited to the maximum number of defect locations per unit volume given by $N^\text{max}_i$.   Given these assumptions, the rate of creation of absorption centres for a defect of type $i$ is given by Equation \ref{eqn_NaRate}.
 
 \begin{equation}
 \label{eqn_NaRate}
 \frac{d N_i(D) }{d D}= \frac{N^\text{max}_i - N_i(D) }{\rho_i}
 \end{equation}
 
  Using Equation \ref{eqn_NaRate} the expression for $A_i(D)$  is given by Equation \ref{eqn_Amin} where $A_i^\text{min}$ is the minimum absorption length reached when all available defects have been converted into absorption centres.  
 
 \begin{equation}
 \label{eqn_Amin}
 A_i (D) = \frac{\kappa}{N_i^\text{max} (1 - e^{-D/  \rho_i} )} = \frac{A_i^\text{min}}{ F(D,\rho_i)}
 \end{equation}
 
\noindent where $F(D,\rho_i) = (1 - e^{-D/  \rho_i} ) $.  From Equation \ref{eqn_Amin}, each defect type is described by two parameters, $\rho$ and $A^\text{min}$.  The parameter $A^\text{min}$ characterizes the number of defect locations in a crystal and the parameter $\rho$ is the parameter characterizing the radiation hardness associated with that defect type.  Substituting Equation \ref{eqn_Amin} into Equation \ref{eqn_rel_LY} for two defect types ($a$ and $b$) the relative light output loss as a function of dose is given by Equation \ref{eqn_model} where $O^\text{min}=L/A^\text{min}$, the optical density.
 
 \begin{equation}
 \label{eqn_model}
  \begin{split}
 \text{LY}(D) &= \exp \Big(-L \Big[ \frac{F(D,\rho_a)}{A^\text{min}_a} +  \frac{F(D,\rho_b)}{A^\text{min}_b}  \Big]  \Big) \\
  &=  \exp \Big(- \Big[ O^\text{min}_a F(D,\rho_a) +   O^\text{min}_b F(D,\rho_b)  \Big]  \Big) \\                    
     \end{split}
 \end{equation}

\begin{figure}[H]
\centering

\subfloat[]{\includegraphics[width=0.9\textwidth]{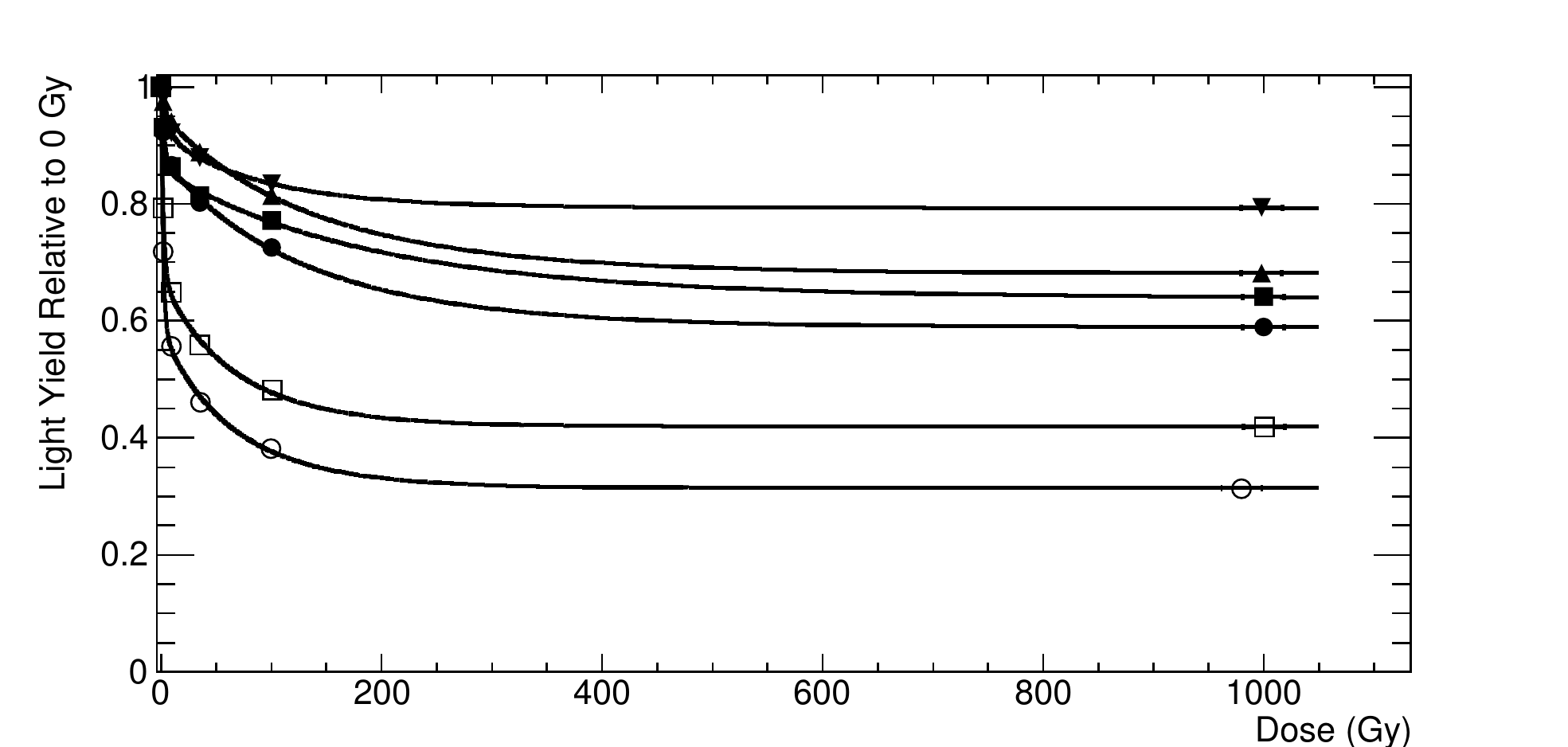}\label{fig:model_LargeDose} } 

\subfloat[]{\includegraphics[width=0.9\textwidth]{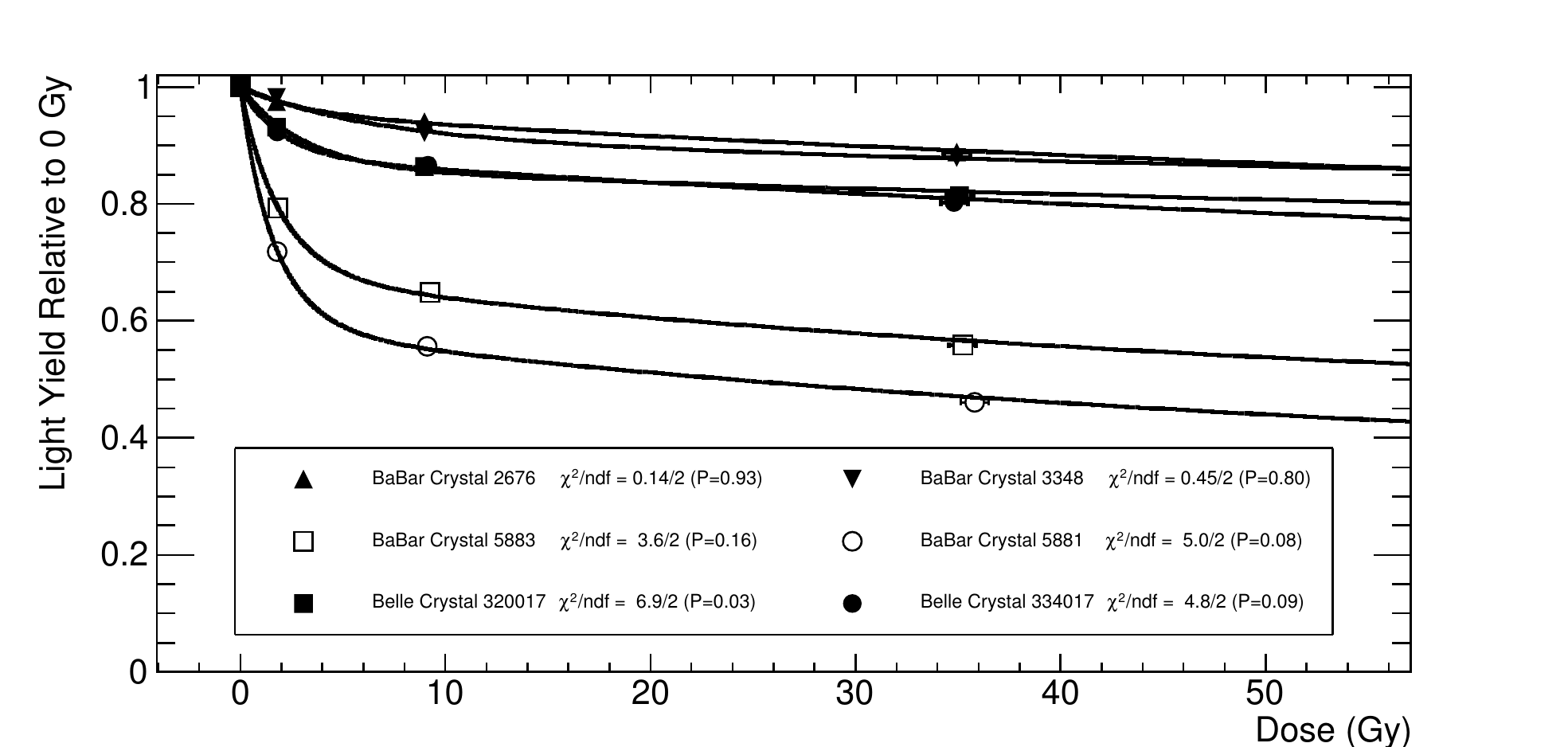}\label{fig:model_SmallDose} } 

\caption{ Model fits to \babar{} and Belle crystals. a) Large Dose. b) Small Dose. } 
\label{fig:modelfitindependent} 
\end{figure}

 \subsection{Model Fits to Experimental Data}
\label{ModelFitsSection}
 
Using Equation \ref{eqn_model} the data from Figure \ref{fig:AllLyvsDose} was fit independently for each crystal.  Fit results are shown in Figure \ref{fig:modelfitindependent} demonstrating that Equation \ref{eqn_model} accurately describes the plateauing trend observed in the light output loss for all crystals.

\begin{figure}[H]
\centering

\subfloat[]{\includegraphics[width=0.75\textwidth]{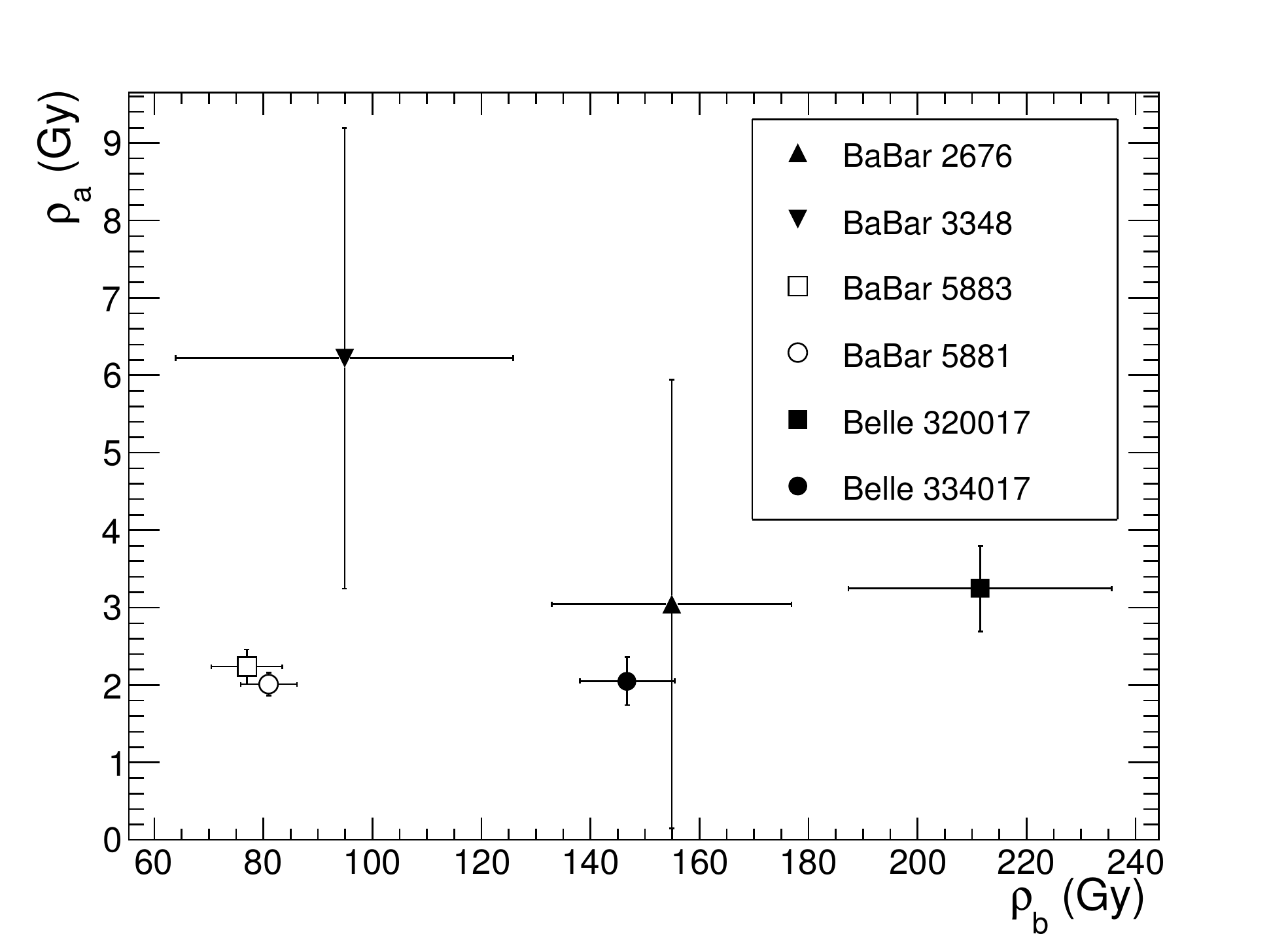}\label{fig:model_corrho} } 

\subfloat[]{\includegraphics[width=0.75\textwidth]{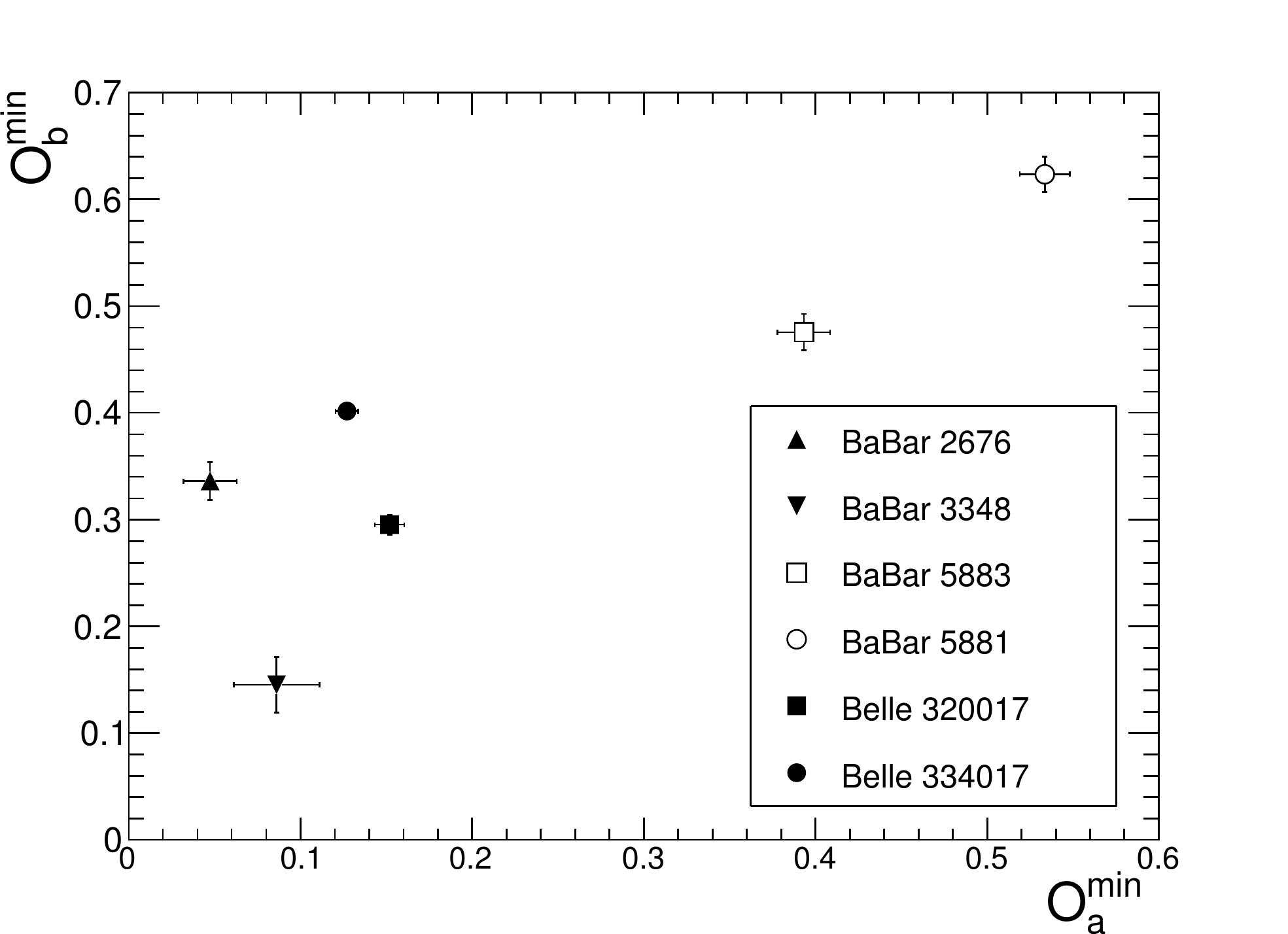}\label{fig:model_corod} } 

\caption{Fit results for a)  $\rho_a$ vs  $\rho_b$. b) $O^\text{min}_b$ vs  $O^\text{min}_a$.} 
\label{fig:modelcor} 
\end{figure}

The correlation between parameters $\rho_a$ and  $\rho_b$ are shown in Figure \ref{fig:model_corrho}.  These parameters are properties of the defects and should be similar between crystals with identical defects.  The average value for $\rho_a$ is calculated to be $2.1\pm0.1$ Gy with a $\chi^2$/ndf of 7/5 (Probability = $22\%$) suggesting this impurity is common to all crystals.  For the second defect parameter $\rho_b$, all Shanghai Institute of Ceramics crystals having approximately $\rho_b = 160$ Gy and Kharkov Institute for Single Crystals having approximately $\rho_b = 80$ Gy  suggesting this defect is manufacturer dependent.  

The correlation between model parameters $O^\text{min}_a$ and  $O^\text{min}_b$ is plotted in Figure \ref{fig:model_corod}.  We note that there appears to be a linear correlation between $O^\text{min}_a$ and  $O^\text{min}_b$ for crystals 5883, 5881 and 3348. This model may be used to predict the light output loss of crystals in the Belle II calorimeter as the dose increases with time from SuperKEKB operations.

\section{Conclusions}

In anticipation of large beam background doses expected to be present in the Belle II calorimeter during the operation of SuperKEKB, the radiation hardness of several 30 cm long CsI(Tl) crystals is measured up to 1000 Gy, a factor of 10 more than the estimate of the 10 year accumulated dose in the Belle II calorimeter.   Measurements of the relative light yield, longitudinal light yield non-uniformity, energy resolution and timing resolution are presented at doses of 2, 10, 35, 100 and 1000 Gy. Up to 1000 Gy, the light yield of 30 cm long CsI(Tl) crystals will require an order of magnitude more dose to have the same relative drop in light yield.   A comparison of non-uniform and uniform irradiation methods demonstrated identical degradation of crystal light output loss and longitudinal light uniformity.  Using our measurements of the light output loss observed in the Belle crystals at 2 Gy it was shown that uniform doses from \co{} reproduce beam background doses at $e^+e^-$ colliders. The plateauing behaviour in the light output loss is well described by a model consisting of two defect types in the crystals creating absorption centres reducing the absorption length of the scintillation photons.

\acknowledgments

The authors would like to thank: 

The \babar{} and Belle collaborations for supplying crystal samples to study and B. Downton, E. Mainegra-Hing and M. McEwen at Measurement Science and Standards, National Research Council of Canada.


\end{document}